\documentstyle[12pt,epsfig]{article}
\def\lsim{\mathrel{\raise3pt\hbox to 8pt{\raise -6pt\hbox{$\sim$}\hss{$<$}}}}
\newcommand{\boldpi}{\mbox{\boldmath $\pi$}}
\newcommand{\boldpip}{\mbox{\boldmath $\pi^{\prime}$}}
\newcommand{\boldtau}{\mbox{\boldmath $\tau$}}

\newcommand{\boldeps}{\mbox{\boldmath $\epsilon$}}
\newcommand{\vecq}{\vec{q}}
\newcommand{\vecqp}{{\vec{q}}^{\; \prime}}
\newcommand{\fpi}{f_{\pi}}
\newcommand{\mpi}{m_{\pi}}

\newskip\humongous \humongous=0pt plus 1000pt minus 1000pt
\def\caja{\mathsurround=0pt}

\newif\ifdtup
\def\panorama{\global\dtuptrue \openup1\jot \caja
        \everycr{\noalign{\ifdtup \global\dtupfalse
        \vskip-\lineskiplimit \vskip\normallineskiplimit
        \else \penalty\interdisplaylinepenalty \fi}}}
\def\eqalignno#1{\panorama \tabskip=\humongous
        \halign to\displaywidth{\hfil$\displaystyle{##}$
        \tabskip=0pt&$\displaystyle{{}##}$\hfil
        \tabskip=\humongous&\llap{$##$}\tabskip=0pt
        \crcr#1\crcr}}
\begin{document}
\vspace*{-0.6in}
\hfill \fbox{\parbox[t]{1.12in}{LA-UR-98-2346 KRL MAP-229}}\hspace*{0.35in}
\vspace*{0.6in}

\begin{center}

{\Large {\bf Chiral Symmetry and Three-Nucleon Forces}}\\

\vspace*{0.20in}
by\\
\vspace*{0.20in}
J.\ L.\ Friar and D.\ H\"uber\\
Theoretical Division \\
Los Alamos National Laboratory \\
Los Alamos, NM  87545 \\
\vspace*{0.20in}
and\\
\vspace*{0.20in}
U.\ van Kolck \\
Kellogg Radiation Laboratory, 106-38\\
California Institute of Technology \\
Pasadena, CA 91125 \\
\end{center}
\vspace*{0.25in}

%\pagebreak
\begin{abstract}
After a brief review of the role three-nucleon forces play in the few-nucleon
systems, the chiral-perturbation-theory approach to these forces is discussed. 
Construction of the (nominal) leading- and subleading-order Born terms and
pion-rescattering graphs contributing to two-pion-exchange three-nucleon forces
is reviewed, and comparisons are made of the types of such forces that are used
today.  It is demonstrated that the short-range $c$-term of the Tucson-Melbourne
force is unnatural in terms of power counting and should be dropped.  The class
of two-pion-exchange three-nucleon forces then becomes rather uniform.
\end{abstract}

\pagebreak
\begin{center}
{\large {\bf Introduction}}\\
\end{center}

       Three-nucleon ($3N$) forces have come under increasing scrutiny recently
\cite{hf}. Although these forces are rather weak, they are playing an important
role in the theory of few-nucleon systems, where computational advances permit
calculation of new observables that are challenged by experiment
\cite{report,joe}.  The most-recent (second-generation) nucleon-nucleon ($NN$)
potentials \cite{pots,AV18} fit the entire $NN$ data base rather well (rivaling
phenomenological partial-wave analyses in the best cases) and lead to
predictions for most $3N$ observables that are in good agreement with
experiment.  In a few cases, such as the $A_y$ puzzle \cite{hf,report} and the
binding energies of few-nucleon ground states \cite{li6}, there are inadequacies
with this methodology that have focused attention on three-nucleon forces.

       All realistic $NN$ forces underbind the triton \cite{2nd}, and small
differences among them can be traced to nonlocalities.  Three-nucleon forces are
incorporated into the Hamiltonian and adjusted to achieve the correct triton
binding.  With this addition $^4$He is properly bound \cite{he4}, while the two
$^5$He p-levels have a splitting roughly 30\% too small \cite{he5}. Binding of A
= 6-8 ground and low-lying excited states is too low \cite{li6}.

       The best-studied of these problems is the $A_y$ puzzle.  The calculated
asymmetry ($A_y$) in neutron-deuteron and proton-deuteron scattering at low
energies is 25-30\% too small, which looks suspiciously similar to the $^5$He
problem, since $A_y$ is most sensitive to spin-orbit forces.  A recent analysis
of the former problem concludes \cite{hf} that reasonable changes in the $NN$
force will not resolve the puzzle and that one should implement refined $3N$
force models.  Although credible examples of these models first began to appear
40 years ago \cite{fm}, technical problems associated with nuclear-force
construction hampered the effort, and general acceptance of such forces was
delayed until it was demonstrated that good $NN$ forces could not reproduce the
triton binding energy.

       Construction of potentials always involves theoretical {\it choices},
since a potential is a subamplitude (an off-shell part of an amplitude) that
when iterated (in the Schr\"odinger equation, for example) produces observables
(on-shell amplitudes or energies).  The off-shell question has always been a
murky one, since it is usually ill defined. Nevertheless, the {\it same}
Lagrangian (i.e., the same theory) can lead to different potentials, although
they should individually produce identical observables.  Coupled to this is the
worse problem of unraveling the underlying strong-interaction physics (i.e.,
deciding on a Lagrangian or equivalent formalism to use).  In the early days a
frequently asked question \cite{tm} was: how does one account for the off-shell
nature of (virtual) pions exchanged between nucleons?  Faced with such daunting
theoretical obstacles, all models were simplified.  Nonlocality
(nucleon-momentum dependence) was typically ignored, for example.  The early
history of the field is well reviewed in Refs. \cite{rome,font}.

       Since these early beginnings a new formalism \cite{cpt,bern,cptnuc,bkm}
has been developed for implementing strong-interaction physics in low-momentum
(for nucleons) regimes:  chiral perturbation theory (CPT).  This technique
implements (approximate) chiral symmetry (manifested by the quarks in QCD) in
constructing the strong-interaction building blocks, which are then assembled in
{\it all} possible ways in the most general Lagrangian consistent with the
symmetry. At the same time, the entire framework is organized with a
power-counting scheme.  A successful perturbation theory must guarantee that
succeeding orders diminish, and chiral symmetry provides the constraints
mandating that more complex calculations (loops, etc.) should yield
progressively smaller results, even though strong-interaction coupling constants
are not small. This scheme also provides a testing mechanism for nuclear
interactions: naturalness and naive dimensional power counting \cite{ndpc}.

       Chiral perturbation theory simplifies the old-fashioned nuclear-physics
approach of incorporating into a field theory all known meson and baryon
resonances with energies less than some large (arbitrary) cutoff.  All such
heavy resonances (with the possible exception of the low-lying $\Delta$ isobar,
which is ignored here for simplicity) are subsumed in short-range (point-like)
vertices.  In the usual SU(2) approach this means that only pion and nucleon
fields contribute explicitly, although the entire zoo of heavy elementary
particles contributes implicitly to the phenomenological constants of the 
theory.

       Two scales that set the strength of the Lagrangian building blocks are
$f_{\pi} \sim 93$ MeV (the pion-decay constant) and $\Lambda \sim 1$ GeV (the
large-mass QCD scale).  Overall powers of $\Lambda$ must be negative (i.e.,
$\Lambda^{-\Delta}$, with $\Delta \geq 0$), since they arise from the frozen
propagation of the heavy states, and interactions in the Lagrangian are
organized by these powers:  ${\cal L}^{(\Delta)}$. Dimensionful coupling
constants in this scheme can be written as powers of $f_{\pi}$ and $\Lambda$
times dimensionless coupling constants $\sim \pm 1$. The latter requirement is
called naturalness.  ``Unnatural'' implies very small or very large (compared to
1) and of either sign. We will use this test later.

       We wish to examine and compare the two-pion-exchange three-nucleon forces
($3N$Fs) that incorporate at least minimal phenomenology from $\pi$-$N$
scattering.  There are basically four types (plus variants of each that we will
not treat):  (1) Tucson-Melbourne force \cite{tm} (the first of this class), 
based on current-algebra
arguments; (2) Brazilian force \cite{brazil}, based on a chiral Lagrangian and a
supplemental current-algebra constraint; (3) Texas \cite{texas} force, based on
chiral perturbation theory; (4) Ruhr(Pot) force \cite{rp}, based on non-chiral
Lagrangians.  Each contains a $\sigma$-term (or functional equivalent) for
s-wave, isospin-symmetric pions, as well as p-wave pions in both
isospin-symmetric and -antisymmetric configurations, such as might arise from
virtual $\Delta$-isobar excitation. We note that the august Fujita-Miyazawa
\cite{fm} $3N$F contained equivalents of all these elements (although the s-wave
part was dropped) and the Urbana-Argonne \cite{ua} model contains a conventional
Fujita-Miyazawa $\Delta$-mediated force plus an intermediate-range isospin- and
spin-independent component.

\begin{center}
{\large {\bf Chiral Perturbation Theory}}\\
\end{center}

       We will make our comparisons using the framework of CPT, which allows us
to define the theory in a consistent and transparent way.  The relevant parts of
the leading-order Lagrangian (corresponding to $\Delta = 0$), ${\cal L}^{(0)}$,
are given by \cite{texas,lag}
$$
 {\cal L}^{(0)}  = \frac{1}{2}[\dot{\boldpi}^{2}-(\vec{\nabla}\boldpi)^{2}
          -m_{\pi}^{2}\boldpi^{2}] 
   + N^{\dagger}[i\partial_{0}-\frac{1}{4 f_{\pi}^{2}} \boldtau \cdot
         (\boldpi\times\dot{\boldpi})]N +\frac{g_{A}}{2 f_{\pi}} 
 N^{\dagger}\vec{\sigma}\cdot\vec{\nabla}(\boldtau\cdot\boldpi)N \, , \eqno (1)
$$
whose three terms correspond to free pions, the free-nucleon energy and
Weinberg-Tomozawa two-pion interaction, and the usual pion-nucleon interaction. 
We have simplified the nonlinear realizations of the SO(4) symmetry \cite{texas}
and dropped terms that would have added even numbers of pion fields to all terms
with pion fields; we do not require such terms in what follows. In addition, the
$\Delta = 1$ Lagrangian, ${\cal L}^{(1)}$, is given by \cite{texas,lag}
$$\eqalignno{
 &{\cal L}^{(1)} 
        =\frac{1}{2m_{N}}\left [N^{\dagger}\vec{\nabla}^{2}N
        -\frac{1}{4 f_{\pi}^{2}}N^{\dagger}\{\boldtau\cdot
        (\boldpi\times\vec{\nabla}\boldpi),\cdot \, \vec{p}\, \}N 
   +\frac{g_{A}}{2 f_{\pi}}N^{\dagger}\{ \boldtau\cdot\dot{\boldpi},
        \vec{\sigma}\cdot\vec{p}\, \} N \right ]   &\cr
  & +\frac{1}{f_{\pi}^{2}}N^{\dagger}[(c_2 +c_3 - \frac{g_A ^2}{8 m_{N}})
        \dot{\boldpi}^{2} -c_3 (\vec{\nabla}\boldpi)^{2} 
        -2c_1 m_{\pi}^{2} \boldpi^{2} -
        \frac{1}{2} (c_4 + \frac{1}{4m_{N}}) 
        \varepsilon_{ijk} \varepsilon_{abc} \sigma_{k} \tau_{c} 
        \partial_{i}\pi_{a}\partial_{j}\pi_{b}]N &\cr  
  & -\frac{d_1}{\fpi} 
        N^{\dagger}\vec{\sigma}\cdot\vec{\nabla}(\boldtau \cdot \boldpi)N\,
        N^{\dagger}N
        -\frac{d_2}{2 \fpi} \varepsilon_{ijk} \varepsilon_{abc} 
        \partial_{i}\pi_{a}  
        N^{\dagger}\sigma_{j}\tau_{b}N\, N^{\dagger}\sigma_{k}\tau_{c}N 
        +\cdots \, ,             &(2) \cr}
$$
where terms with additional pion fields have been dropped, and we have not
listed \cite{texas} three separate spin- and isospin-dependent short-range $3N$F
terms ($\sim (N^{\dagger} N)^3)$ with coefficients, $e_i$. We have also ignored
isospin violation. Where appropriate we have adopted the notation of Ref.
\cite{bkm} and have explicitly incorporated higher-order terms resulting from a
nonrelativistic reduction of the pseudovector-coupling Born term.  The
phenomenological coefficients $c_i$ and $d_i$ must be determined from 
experiment.

        We have not written down explicit $\Delta$-isobar contributions above.
They are implicitly included in the phenomenological coefficients. This hides
the fact that those coefficients that contain tree-level $\Delta$ contributions
are expected to be larger than ones that do not by a $\Lambda/(m_\Delta -m_N)$
factor. The alternative is to include a $\Delta$ field and count it as a nucleon
field\cite{texas}. This shifts the nominal order of the isobar effects but of 
course not their numerical value, and it unnecessarily complicates the following
discussion.

      For later use we also list infinitesimal generators for the (approximate)
axial symmetry present in this Lagrangian, where again we ignore terms with more
than two pion fields:
$$\eqalignno{
\boldpi & \rightarrow \boldpi -\fpi \boldeps\, , &(3a)\cr
N &\rightarrow N - i \frac{\boldeps \cdot \boldtau \times \boldpi}{4 \fpi} N
\, , &(3b)\cr}
$$
where $\boldeps$ is (a constant) infinitesimal.  Under this transformation the
three terms in Eq. (1) are separately invariant in the limit of vanishing pion
mass, as are the first bracketed term and each remaining term in ${\cal
L}^{(1)}$ (in the same limit).  Thus, the Lagrangian in Eqs. (1) and (2) is
term-by-term (as we have written them) invariant, except for the pion-mass and
$c_1$-term (also conventionally known as the $\sigma$-term): $-4\, \mpi^2 \, c_1
= \sigma$.

   It is important to note that the Lagrangians ${\cal L}^{(i)}$ are not unique.
Redefinition of the (unphysical) fields leads to other forms. The form we have 
chosen satisfies chiral constraints in a term-by-term fashion, rather than 
relying on cancellations between sets of terms. It is only important that the
chosen form have sufficient generality (i.e., enough linearly independent 
terms). Different forms will then be physically equivalent on shell, but will 
in general be different off shell. Off-shell differences do not affect physical
processes. Note that the Lagrangian of Ref. \cite{cf}, which is based on a 
non-relativistic reduction of the relativistic pseudo-vector pion-nucleon 
coupling, used an off-shell extension specified by a continuous parameter, 
$\mu$. Only the choice $\mu=1$ corresponds to Eq. (2) and only that choice 
satisfies term-by-term chiral symmetry. Amplitudes calculated using various 
values of $\mu$ correspond to a unitary transformation of the Hamiltonian and 
therefore do not alter physical amplitudes (although they are different off 
shell). We note\cite{cf} that many of the older papers in the field have 
implicitly adopted different values of $\mu$ [viz., -1,0,1].

\begin{figure}[htb]
  \epsfig{file=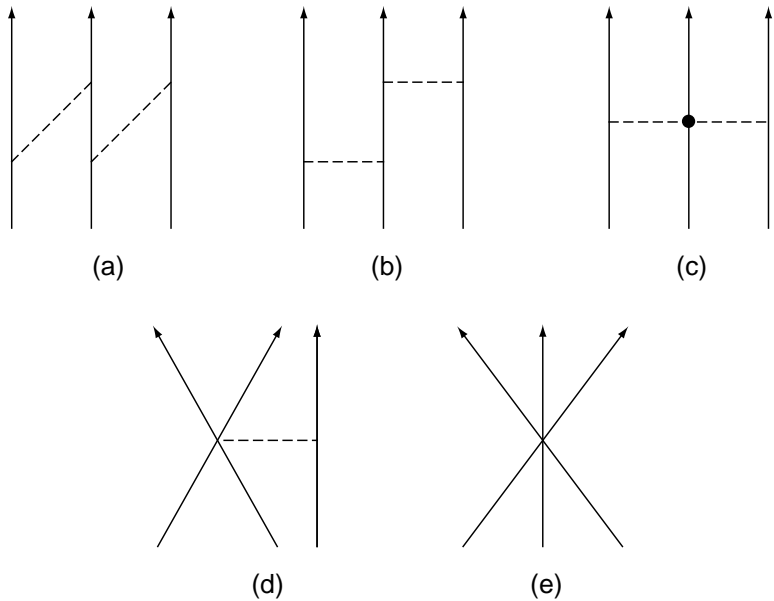,height=3.0in,bbllx=-75pt,bblly=210pt,bburx=600pt,
bbury=600pt}
  \caption{Various three-nucleon-force components that arise in subleading order
in chiral perturbation theory, as discussed in the text.}
\end{figure}

      In order to determine the $3N$F to (nominal) subleading order, we need to
calculate the diagrams of Fig. (1).  The two interaction terms in ${\cal
L}^{(0)}$ together with the first two terms in ${\cal L}^{(1)}$ are usually
called relativistic Born terms, and are separately calculated using (the many
orderings of) Figs. (1a) and (1c), and then subtracting the iteration of the
one-pion-exchange potential (OPEP) given in (1b).  In the static (leading-order)
limit $(m_N \rightarrow \infty)$ they have long been known to vanish
\cite{yg,cf}. If one works to subleading order one is faced with choices, 
because different off-shell choices for the subtracted OPEP lead to different
forms for the $3N$F.  Thus, the {\it choice} of form for OPEP (to order
$(v/c)^2$) determines the form of this (Born-term) part of the $3N$F. The reader
is referred to Refs. \cite{texas} and \cite{cf}, where different off-shell
choices are made.  The complete $(\mu,\nu)$ off-shell ambiguity is discussed in
the latter reference and approximate Lorentz invariance is demonstrated. The 
former ambiguity arises from a nucleon-field transformation (a ``chiral 
rotation'') that breaks term-by-term chiral invariance, as we discussed below 
Eq. (2). Different values of $\mu$ have been implicitly assumed in the past by 
differing treatments of the 
Born terms (see Appendix of Ref. \cite{cf}). The $\nu$-dependence arises through
differing treatments of the difference between (four-vector) $q^2$ and 
$\vecq^{\, 2}$ (see Eq. (4b) below), and is sometimes called the quasipotential
parameter. Different quasipotential equations correspond (in part) to different 
values of $\nu$, and the values [0,1/2,1] have been commonly used \cite{cf}.
Different values of $\nu$ correspond to different off-shell amplitudes, but 
unitarily-equivalent on-shell values.
Other calculations have ignored part or all of the subleading-order
Born-term contributions. We will ignore the Born terms in what follows.

       The remaining 9 terms of ${\cal L}^{(1)}$ [labeled by $c_i, d_i, e_i$]
generate $3N$Fs of the type in Fig. (1c) [$c_1, c_3, c_4$], Fig. (1d) [$d_1,
d_2$], and Fig. (1e) [$e_1, e_2, e_3$].  The $\dot{\boldpi}^2$ term in ${\cal
L}^{(1)}$ generates contributions of $\Delta = 3$ size (each time derivative is
the same as a nucleon-energy difference) and can be neglected.  A wide range of
physics is subsumed in each category.  The $c_3$ and $c_4$ terms receive
important contributions from $\Delta$-isobars at the blob of Fig. (1c), while a
heavy scalar-isoscalar meson would likewise contribute to $c_1$.  We note that
all of the models we will compare contain this important physics, either through
phenomenological input or via explicit heavy-particle intermediate states.

       We summarize by noting that the Born term from ${\cal L}^{(0)}$, the
$c_i$ $\pi$-rescattering terms, the $d_i$ one-pion-exchange terms, and the
(purely) short-range $e_i$ terms are all nominally the same size, although large
$\Delta$-isobar contributions can be expected to make some of the terms larger
than others.  We will not discuss the $d_i$ and $e_i$ terms further. This force
was first derived in Ref. \cite{texas}.

\begin{center}
{\large {\bf Comparisons}}\\
\end{center}

       To facilitate comparisons we adopt the familiar framework of the
Tucson-Melbourne collaboration \cite{tm} for the Born-subtracted amplitudes
\cite{cp}
$$
S = 1 - i \, T \, , \eqno (4a)
$$
$$V_{3NF} = T = \left (\frac{g_A}{2 \fpi} \right )^2 \frac{ \vec{\sigma}_1 \cdot
\vecq \; \vec{\sigma}_2 \cdot \vecqp} {(\vecq^{\; 2} + \mpi^2)
(\vec{q}^{\; \prime \; 2} +\mpi^2)} 
\left [ - F^{\alpha \beta}\; \tau_1^\alpha \tau_2^\beta \right ]\, , \eqno (4b)
$$
$$
t^{\alpha \beta}_{\pi N} = -F^{\alpha \beta} \cong \delta^{\alpha \beta} \, 
\left [
 a + b \, \vecq \cdot \vecqp +c \, (\vecq^{\; 2} + \vecq^{\; \prime \;  2} )
 \right ] -d\,( \tau_3^{\gamma} \epsilon^{\alpha \beta \gamma}\; \vec{\sigma_3} 
 \cdot \vecq \times \vecqp ) \; , \eqno (4c)
$$
where $\delta$-functions, phase-space factors, etc., have been ignored, and the 
invariant amplitudes of \cite{tm,cp} have been expanded in $1/m_N$.

\begin{figure}[htb]
  \epsfig{file=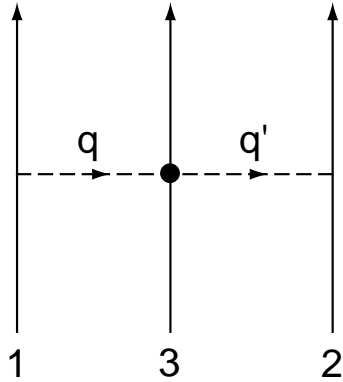,height=2.0in,bbllx=-200pt,bblly=-25pt,bburx=150pt,
bbury=150pt}
  \caption{Contribution to the three-nucleon force arising from a pion emitted
  by nucleon 1 and scattering from nucleon 3 before being absorbed by nucleon 
  2.}
  \end{figure}
Equation (4b) is illustrated in Fig. (2), showing nucleon (3) scattering a pion
emitted by nucleon (1) and absorbed by nucleon (2). The T-matrix for $\pi$-$N$
scattering (alone) is denoted $t^{\alpha \beta}_{\pi N}$ and is usually
rewritten in terms of $F^{\alpha \beta}$, where $\alpha$ and $\beta$ are the
isospin labels of the initial and final pions.  The final expression in Eq. (4c)
holds for pions that have a low momentum ($\lsim \mpi$). Summing over the 
symmetric permutations of (1,2,3) in Fig. (2) leads to the complete 
three-nucleon potential.

One easily finds from Eq. (2):
$$
F^{\alpha \beta} = \frac{ \delta^{\alpha \beta}}{\fpi^2} \, \left [ 2 \omega 
\omega^{\prime} (c_2 + c_3) -2 c_3 \, \vecq \cdot 
\vecqp -4 c_1 \mpi^2 \right ] - \frac{
\tau_3^{\gamma} \epsilon^{\alpha \beta \gamma}\; \vec{\sigma_3} 
 \cdot \vecq \times \vecqp}{\fpi^2} [c_4] \, , \eqno(5)
$$
where $\omega$ and $\omega^{\prime}$ are the initial and final pion energies. We
have dropped Born term contributions to $c_2$ and $c_4$ in accordance with our
earlier discussion. Eq. (5) together with the dropped pieces generates the CPT
$\pi$-$N$ amplitude to $O(Q^2)$. Calculations including loops and new parameters
at  $O(Q^3)$ have also been performed. They have been used with different pieces
of $\pi$-$N$ scattering data to determine the coefficients $c_i$. In table (1)
we list some of these determinations. Earlier fits \cite{bkm2,bkm,bkm3,moj} were
made to different sets of threshold and sub-threshold parameters obtained from
dispersion analyses of older data. Newer fits \cite{fms} were made to different
phase-shift analyses (PSAs), the last two in Table (1) including the more modern
meson-factory data. The $O(Q^3)$ determinations are consistent with each other
when their error bars (not shown) are considered, except for $c_1$, which
reflects the higher value for the $\sigma$-term in the newer PSAs. Note that the
coefficients $c_2$, $c_3$, and $c_4$, which receive contributions from the
$\Delta$ at tree level, are larger than $c_1$, as expected\cite{rob}. 

\begin{table}[htb]
\centering
\caption{Low-energy CPT coefficients in GeV$^{-1}$ from several recent fits.}

\hspace{0.25in}

\begin{tabular}{|l||cccc|}
\hline
{Fit}\rule{0in}{2.5ex} & $c_1$ & $c_2$ & $c_3$ & $c_4$\\ 
\hline \hline
$O(Q^2)$\cite{bkm2}\rule{0in}{2.5ex}   & -0.64 &  1.78 & -3.90 &  2.25\\
$O(Q^3)$\cite{bkm}                     & -0.87 &  3.30 & -5.25 &  4.12\\ 
$O(Q^3)$\cite{bkm3}                    & -0.93 &  3.34 & -5.29 &  3.63\\ 
$O(Q^3)$\cite{moj}                     & -1.06 &  3.40 & -5.54 &  3.25\\
$O(Q^3)$\cite{fms}                     & -1.27 &  3.23 & -5.93 &  3.44\\ 
$O(Q^3)$\cite{fms}                     & -1.47 &  3.21 & -6.00 &  3.52\\
$O(Q^3)$\cite{fms}                     & -1.53 &  3.22 & -6.19 &  3.51\\ 
\hline
\end{tabular}
\end{table}

{}From the definition (4c) of the ($a,b,c,d$) coefficients we obtain,
$$\eqalignno{
a &= \frac{4\, \mpi^2\, c_1}{\fpi^2} = - \frac{\sigma}{\fpi^2}\, , &(6a)\cr
b &= \frac{2\, c_3}{\fpi^2} \, , &(6b)\cr
c &= \, 0 \, , &(6c)\cr
d &= -\frac{c_4}{\fpi^2}\, . &(6d) \cr}
$$
Note that there is no $c$-term, and that the $a$-term is opposite in sign to the
TM result, although with $c_3 < 0$ and $c_4 > 0$, $b$ and $d$ are negative and 
agree with the corresponding TM signs.  A similar result was found in the 
first of the Brazil-force papers \cite{brazil}, where a field-theoretic 
calculation of isobar contributions was performed. The $\sigma$-term was not 
calculated using Feynman rules derived consistently from a Lagrangian, but 
inferred from a $\pi$-$N$ amplitude derived elsewhere. In a later paper, a 
different off-shell amplitude (the one used in the TM calculation) was 
incorporated. Values of the $a-d$ coefficients for popular three-nucleon 
force models are displayed in Table (2). Note that $a=a^{\prime}+ 2\, \mpi^2\, 
c = 1.03/\mpi$ for the TM force.

\begin{table}[htb]
\centering
\caption{Low-energy pion-nucleon scattering parameters (with Z-graph [Born]
terms removed) for a variety of 2$\pi$-exchange three-nucleon forces We have
also defined $a^{\prime}=a - 2\, \mpi^2\, c$. The quantities $a$ and
$a^{\prime}$ are in units of $\mpi^{-1}$, while $b$, $c$, and $d$ are in units 
of $\mpi^{-3}$.}

\hspace{0.25in}

\begin{tabular}{|l||cccc|}
\hline
{Three-Nucleon Force}\rule{0in}{2.5ex} &$a^\prime$&$b$ & $c$ & $d$\\ 
\hline \hline
Fujita-Miyazawa\cite{fm}\rule{0in}{2.5ex}   &  0.0  & -1.15 & 0.0  & -0.29\\
Tucson-Melbourne\cite{tm,mc}                & -1.03 & -2.62 & 1.03 & -0.60\\ 
Brazil\cite{brazil,mc}                      & -1.05 & -2.29 & 1.05 & -0.77\\ 
Urbana-Argonne\cite{ua,li6}                 &  0.0  & -1.20 & 0.0  & -0.30\\ 
Texas\cite{texas,fms}                       & -1.87 & -3.82 & 0.0  & -1.12\\
Ruhr(Pot)\cite{rp}                          & -0.51 & -1.82 & 0.0  & -0.48\\ 
\hline
\end{tabular}
\end{table}
 
        Given that CPT is a comprehensive approach to calculating
strong-interaction physics based on chiral symmetry and subsumes current
algebra\cite{bkm,tw}, how can the CPT (corresponding to derivatively-coupled 
pions) amplitude [Eq. (5)] and TM amplitude [Eq. (4c)] differ?

       We answer that question by noticing that the difference resides only in
terms that vanish when the pions are on-shell (as we shall see in Eq. (11)). We 
return to the earlier
off-shell discussion, and follow closely the approach of Ref. \cite{tw}.
Off-shell amplitudes are not unique, and in a field-theoretic calculation they
depend on the fields chosen to represent pions and nucleons.  Our form was
chosen to satisfy chiral symmetry term-by-term, thereby attaining manifest power
counting. Current-algebra constraints at certain off-mass-shell points \cite{cp}
are not satisfied by our isospin-even $\pi$-$N$ amplitude, $F^{(+)}$ $[F^{\alpha
\beta} = \delta^{\alpha \beta} F^{(+)} + \cdots]$.  These points all correspond
to vanishing (four-vector) $q \cdot q^{\prime}$, as well as $\omega$ and
$\omega^{\prime}$ (to the order we work). Consequently, we can ignore the $c_2$-
and $c_3$-terms in Eq. (5) and concentrate on the remaining term, which can be
written in the form
$$
F^{({+})}_{\rm CPT} = \frac{\sigma}{\fpi^2}\, , \eqno (7)
$$
which holds everywhere.

        Again following Ref. \cite{tw}, we redefine the pion field as
$$
\boldpi^{\prime} = \boldpi\, (1-\frac{\sigma}{\mpi^2 \fpi^2}\, N^{\dagger} N )
\, , \eqno (8)
$$
and work only to order $\Delta = 1$ (since $\sigma \sim 1/ \Lambda$).
Substituting Eq. (8) into ${\cal L}^{(0)}$, we generate the extra terms
$$
\Delta {\cal L}^{(1)}  = -\frac{\sigma \, }{\mpi^2 \fpi^2}
\left [ N^{\dagger} N (\boldpip \cdot  \, \Box \boldpip + \mpi^2 
\boldpi^{\prime \; 2}) -\frac{g_A}{2\, \fpi}
N^{\dagger} \, \vec{\sigma} \cdot \vec{\nabla} [\boldtau \cdot \boldpip
N^{\dagger} N ]\, N\, \right ] + \cdots \, , \eqno (9)
$$
The last term involves four nucleon fields and is not immediately required. The
three terms in ${\cal L}^{(1)}$ and $\Delta {\cal L}^{(1)}$ involving $\sigma$
and two nucleon fields lead to
$$
F^{(+)}_{\rm CA} = \frac{\sigma}{\mpi^2 \fpi^2}\, 
( q^{\; 2} + q^{\; \prime \; 2} - \mpi^2)\, , \eqno (10)
$$
which agrees with Eq. (7) at any on-shell (e.g., Cheng-Dashen) point $(q^2 =
q^{\prime \, 2} = \mpi^2)$, but vanishes at the Adler points $(q^2 = \mpi^2,
q^{\prime \, 2} = 0$, and $q^{\prime \, 2 }= \mpi^2, q^2 = 0)$, and at the
Weinberg point $(q^2 = q^{\prime \, 2} = 0)$ has the value: $F^{(+)}_{\rm CA} 
= -\sigma
/\fpi^2$.  Equation (10) therefore agrees with the usual current-algebra
constraints \cite{tm,cp}, as does our entire amplitude, $F^{(+)}$, in the new
pion-field basis. Thus, there is no conflict here between CPT (with 
derivatively-coupled pions) and an approach based on current algebra. The
only difference is in the {\it choice} of fields used to specify the chiral 
Lagrangian, and observables calculated for physical processes must be identical.

       In the TM approach it was noted that rewriting Eq. (10) in terms of {\it
inverse} pion propagators,
$$
F^{(+)}_{\rm CA} = \frac{\sigma}{\fpi^2}\,  + \frac{\sigma}{\mpi^2 \fpi^2}\, 
( q^{\; 2} - \mpi^2 + q^{\; \prime \; 2} -\mpi^2 )\, , \eqno (11)
$$
allows cancellation of the pion propagators in Fig. (2). The first (constant)
term reproduces Eq. (7) ($F^{(+)}_{\rm CPT}$). This rearrangement amounts to 
undoing the field transformation in Eq. (8) that led to Eq. (9), and leads to 
an effective $a$-term ($a^{\prime}$) that has a common sign for all models: 
$a^{\prime}
= a - 2\, \mpi^2 c$.  Cancelling the inverse propagators in the second term in 
Eq. (11) leads to a new short-range-plus-pion-range $3N$F:
$$
-\left (\frac{g_A}{2 \fpi} \right )^2 
\frac{\sigma}{\mpi^2 \fpi^2}
\vec{\sigma}_1 \cdot \vecq \; \vec{\sigma}_2 \cdot \vecqp
(\frac{1}{\vecq^{\; 2} + \mpi^2}+
\frac{1}{\vec{q}^{\; \prime \; 2} +\mpi^2})\boldtau_1 \cdot \boldtau_2 \, . 
\eqno (12)
$$
However, a three-nucleon force of the same type is generated by the last term 
in Eq. (9), comprised of four nucleon fields and one pion field, together with 
the last term in Eq. (1): two graphs as in Fig. (1d) give
$$
\left (\frac{g_A}{2 \fpi} \right )^2 
\frac{\sigma}{\mpi^2 \fpi^2}
\vec{\sigma}_1 \cdot \vecq \; \vec{\sigma}_2 \cdot \vecqp
(\frac{1}{\vecq^{\; 2} + \mpi^2}+
\frac{1}{\vec{q}^{\; \prime \; 2} +\mpi^2})\boldtau_1 \cdot \boldtau_2 \, . 
\eqno (13)
$$
This is {\it exactly equal in size and opposite in sign} to the new short-range 
contribution from the off-shell extrapolation of the $\pi$-$N$ amplitude, Eq. 
(12). This cancellation is to be expected, since our original (chiral) 
Lagrangian produced no such terms to start with, and we have just been 
rearranging terms since then. In summary, the TM approach used a 
current-algebra representation of the amplitude, performed an implicit field 
redefinition to our (CPT) choice of fields, which resulted in an extra 
short-range term in their result. Why did they have an extra term and we do not?

       The TM calculation was predicated upon current-algebra constraints on 
the off-shell $\pi$-$N$ scattering amplitude, which can be reproduced in the
CPT approach, as well, as we have demonstrated. It is not enough, however, to 
worry about just that scattering amplitude, if one constructs a $3N$F. To 
incorporate all of the chiral constraints into the three-nucleon force, 
current-algebra constraints on the pion-production amplitude from two nucleons 
would also be necessary (leading to the last term in Eq. (9)): a daunting task
in the current-algebra approach of TM, but one that is unnecessary in our 
approach. We emphasize that a detailed analysis of the off-shell region of
$\pi$-$N$ scattering (for example) is equivalent to a particular choice of 
fields, and (while interesting) is not necessary for constructing a $3N$F.

If one uses the pion-field redefinition in the symmetry
generators, Eq. (3), one finds that the entire Lagrangian maintains its
original symmetry, but that $\Delta {\cal L}^{(1)}$ generates new
non-invariant terms that cancel against additional contributions from
${\cal L}^{(0)}$ (via the new term 
$\boldeps \sigma N^{\dagger}N/\mpi^2 \fpi$  in the pion generator).  One 
might presume that since all of these terms violate chiral symmetry this poses 
no problem. Unfortunately, chiral-symmetry-breaking terms must vanish in the 
chiral limit. The additional terms in Eq. (9) (being just a redefinition of 
fields) exactly cancel each other in any on-shell amplitude. Individually, the
two terms do not vanish in this limit because the presence of the $1/\mpi^2$ in
Eq. (9) removes the implicit $\mpi^2$ in $\sigma$, and $\sigma/\mpi^2$ does not
vanish in the chiral limit $(\mpi \rightarrow 0)$.  Reiterating, the structure
of the additional terms in Eq. (9) means that they must individually vanish in
that limit, or the entire set of terms must be kept to allow for exact
cancellations between them to restore the proper limit. Because the TM
approach (implicitly) kept only the first term in Eq. (9), that limit could not
be guaranteed for the three-nucleon force. Another way of saying the same thing
is that dimensional power counting (naturalness) is not satisfied for the {\it
individual} terms in Eq. (9).

       One can check this conclusion by dimensional power counting. An 
interaction of the form of the last term in Eq. (9) is chiral-symmetry breaking;
if it alone is to be kept, it has to be implicitly proportional to $\mpi^2$, and
hence is nominally an ${\cal L}^{(3)}$ term. Such ${\cal L}^{(3)}$ coefficients
have a generic size $x \, \mpi^2/\fpi^3 \Lambda^3$, where the dimensionless
coefficient $x$  should be of order 1.  If we equate this to $g_A \sigma/2 \,
\mpi^2 \fpi^3$ (the coefficient of the last term in Eq. (9)), we obtain $x \sim
g_A \sigma \Lambda^3 /( 2 \, \mpi^4) \sim 100$, which is vastly unnatural. The
unnatural coefficient $[g_A \, \sigma / 2\, \mpi^2\, \fpi^3]$
is entirely the result of $\sigma/\mpi^2$ having a finite symmetry limit.

       We recommend that the short-range $c$-term in the TM force be dropped 
(but the full value of $a^{\prime}$ in Table (2) retained; note that the proper
power counting has been maintained in $a^\prime$ by the factor of $\mpi^2$ 
preceding $c$ in the definition of $a^\prime = a - 2 \mpi^2 c$, where now each 
term in this definition vanishes in the chiral limit).
This had been previously advocated by the Brazil group for reasons
having nothing to do with symmetry.  We note that the $d_1$ and $d_2$
terms in the Texas force are also short-range in one pair of nucleons
and of pion range in the other.  These parts of that force (and the 
corresponding terms in the Lagrangian)  satisfy chiral 
constraints, as does a fully short-range force of the generic type contained 
in the UA $3N$F, and shown in Fig. (1e).

        In summary, we have briefly reviewed the class of ``realistic''
three-nucleon forces. We have demonstrated that the short-range $c$-term of
the TM approach is unnatural and should not be kept.

\begin{center}
{\large {\bf Acknowledgments}}\\
\end{center}
The work of JLF and DH was performed under the auspices of the United States 
Department of Energy, while that of UvK was supported under National Science 
Foundation grant PHY 94-20470. The work of D.H. was supported in part by 
the Deutsche Forschungsgemeinschaft under Project No. Hu 746/1-2. We would like 
to thank J. Carlson and M. Mattis of LANL, R. Wiringa and S. Pieper of ANL, 
V. Pandharipande of Univ. of Illinois, J. Adam of TJNAF, J. Eden of Ruhr-Univ. 
Bochum, R. Timmermans of KVI, and S. A. Coon of New Mexico State Univ. for 
their helpful comments and discussion.

\end{document}